\documentclass[reprint,aps,prb,showpacs,amsmath,amssymb,superscriptaddress,nofootinbib,twocolumn]{revtex4-1}

\usepackage{color}
\usepackage{bm}
\usepackage{units}
\usepackage{graphicx}
\usepackage{bbold}
\usepackage{tikz}
\usepackage{graphicx}
\usepackage{subfigure}
\usepackage{bm}
\usetikzlibrary{calc}
\usetikzlibrary{arrows}
\usepackage{hyperref}

\DeclareMathOperator{\im}{Im}
\DeclareMathOperator{\sign}{sign}
\def\normOrd#1{\mathop{:}\nolimits\!#1\!\mathop{:}\nolimits}
\DeclareMathOperator{\Tr}{Tr}
\usepackage{hyperref}
\makeatletter
\newcommand*{\rom}[1]{\expandafter\@slowromancap\romannumeral #1@}
\makeatother

\begin{document}

\title{Skyrmion defects and competing singlet orders in a half-filled antiferromagnetic
Kondo-Heisenberg model on the honeycomb lattice}

\author{Chia-Chuan Liu}
\affiliation{Department of Physics and Astronomy, Rice University, Houston, Texas 77005, USA}

\author{Pallab Goswami}
  \affiliation{Condensed Matter Theory Center and Joint Quantum Institute, Department of Physics, University of Maryland, College Park, Maryland 20742-4111, USA} 
	
	\author{Qimiao Si}
\affiliation{Department of Physics and Astronomy, Rice University, Houston, Texas 77005, USA}

\begin{abstract}
Due to the interaction between topological defects of an order parameter and underlying fermions, the defects can possess induced   fermion numbers, leading to several exotic phenomena of fundamental importance to both condensed matter and high energy physics. One of the intriguing outcome of induced fermion number is the presence of fluctuating competing orders inside the core of topological defect. In this regard, the interaction between fermions and skyrmion excitations of antiferromagnetic phase can have important consequence for understanding the global phase diagrams of many condensed matter systems where antiferromagnetism and several singlet orders compete. We critically investigate the relation between fluctuating competing orders and skyrmion excitations of the antiferromagnetic insulating phase of a half-filled Kondo-Heisenberg model on honeycomb lattice. By combining analytical and numerical methods we obtain exact eigenstates of underlying Dirac fermions in the presence of a single skyrmion configuration, which are used for computing induced chiral charge. Additionally, by employing this nonperturbative eigenbasis we calculate the susceptibilities of different translational symmetry breaking charge, bond and current density wave orders and translational symmetry preserving Kondo singlet formation. Based on the computed susceptibilities we establish spin Peierls and Kondo singlets as dominant competing orders of antiferromagnetism. We show favorable agreement between our findings and field theoretic predictions based on perturbative gradient expansion scheme which crucially relies on adiabatic principle and plane wave eigenstates for Dirac fermions. The methodology developed here can be applied to many other correlated systems supporting competition between spin-triplet and spin-singlet orders in both lower and higher spatial dimensions.    
  
\end{abstract}

\maketitle

\section{Introduction} Competing orders and quantum criticality are two generic features of the rich phase diagrams displayed by  several strongly correlated materials, including heavy fermion systems
\cite{Si-Nature,Coleman-JPCM,Senthiletal,Paschen,Shishido,Schofield,Si_PhysicaB2006,Lohneysen_rmp,SiSteglich}. Of particular significance are the antiferromagnetic phase and competing spin-singlet phases such as charge and bond density waves and unconventional pairings. Therefore, for a comprehensive understanding of the global phase diagrams of many strongly correlated materials, it is essential to gain insights into the relationship among different competing orders, which spontaneously break distinct global symmetries. Within the conventional theme of Landau theory of local order parameters, describing smooth fluctuations or collective modes, order parameters breaking distinct symmetries do not seem to bear any specific relationship. However, the nonperturbative topological defects of order parameters such as 
 domain walls, vortices, skyrmions and hedgehogs can support competing orders as fluctuating objects 
 and thereby contain information about apparently distinct ordered states~\cite{Rajaraman,Shifman2011,Thouless,Polyakov1,Arafune,JackiwRebbi1,THooft,JackiwRebbi2, Callias1,Callias2,Wilczek1,MacKenzie,Kahana,Jaroszewicz,MaNieh,BBoyanovsky,SBoyanovsky,Duncan,Wilczek2, Carena,MurthySachdev,Read,Senthiletal2,Senthiletal3,Sandvik,Kaul,Chalker1,Chalker2,Abanov,Hermele,FisherSenthil,TanakaHu,Saremi,FuSachdev,RoyHerbut,GoswamiSi2,Tsvelik,GoswamiSi1,Grover,ChamonRyu,Herbut1,Moon,Chakravarty1}. 
 In addition, the interaction between fermions and topological defects can be important in strongly correlated electronic systems such as heavy fermion compounds, generically described by effective Kondo-Heisenberg 
 models~\cite{Si-Nature,Coleman-JPCM,Senthiletal,Si_PhysicaB2006,Lohneysen_rmp,SiSteglich,Saremi,GoswamiSi2,Tsvelik,GoswamiSi1,GoswamiSi3}. 
 The strong competition among antiferromagnetism and Kondo singlet formation in addition to spin-singlet superconductivity are essential features 
 of many heavy fermion compounds~\cite{Si-Nature,Coleman-JPCM,Senthiletal,Paschen,Shishido,Schofield,Si_PhysicaB2006,Lohneysen_rmp,SiSteglich},
 and a global phase diagram has been theoretically proposed \cite{Si_PhysicaB2006} 
 which features the transitions between an antiferromagnetic order and a variety of spin-singlet paramagnetic phases.
 This global phase diagram has been studied in the Kondo-Heisenberg models using various microscopic methods \cite{Pixley,Nica}, 
 and has motivated experimental investigations in a number of heavy fermion materials\cite{Custers-2012,Fritsch-2014,Kim-2013,Khalyavin-2013,Mun-2013,Tokiwa-2015,Friedemann-2009,Custers-2010,Luo-2016,Jiao-2015,Tomita2015}.
 However, it remains a theoretical challenge to concretely 
 access the spin-singlet orders (e.g., the heavy fermi liquid phase due to static Kondo singlets) of the paramagnetic phases
starting from the antiferromagnetically ordered side. 
In this work, we are interested in addressing the fluctuating spin-singlet orders supported by gapped skyrmion excitations inside an antiferromagnetically ordered phase of a Kondo-Heisenberg model. We are also interested in identifying the most dominant singlet orders which can be nucleated when the antiferromagnet order is destroyed by quantum fluctuations, causing the collapse of skyrmion excitation gap inside the paramagnetic phase. 

The general problem of interaction between fermions and topological defects is often intractable. 
But valuable insights can be gained by studying specific toy models where fermionic degrees of freedom are modeled by Dirac fermions. 
In this regard, a Kondo-Heisenberg model defined on the honeycomb lattice plays a very instructive role, as the coupling between 
Dirac fermions and antiferromagnetic order parameter can be addressed employing diverse analytical and numerical methods~\cite{Saremi,GoswamiSi2,GoswamiSi3}. In one of our previous work, we have addressed the interaction between Dirac fermions 
and topologically nontrivial skyrmion configuration of antiferromagnetic order parameter, 
by employing perturbative gradient expansion scheme~\cite{GoswamiSi2}. Within such scheme the calculations of triangle diagram 
for Goldstone-Wilczek current are controlled by the inverse of Dirac mass (caused by uniform amplitude of antiferromagnetic order) 
and rely upon adiabatic principle. 

\subsection{Competition between spin Peierls and antiferromagnetic orders}

The simplest situation involves a doublet (two inequivalent valleys or nodes) of spinful Dirac fermions coupled to antiferromagnetic order that simultaneously breaks time reversal and spatial inversion symmetries. The corresponding low energy theory can be described by the effective action
\begin{eqnarray}\label{S1}
S_1=\int d^2x d\tau \bar{\psi}[\gamma_\mu \partial_\mu + g_{\psi} \mathbb{1} \otimes \boldsymbol \eta \cdot \mathbf{n}]\psi,  
\end{eqnarray}
where $\psi$ is a eight-component spinor (incorporating two sublattice, two nodal and two spin degrees of freedom), $\gamma_\mu$ are three mutually anticommuting $4\times 4$ Hermitian matrices operating on sublattice and valley indices, $\mathbb{1}$ is $4\times 4$ identity matrix that operates on sublattice and valley indices, and Pauli matrices $\boldsymbol \eta$ act on spin components. The coupling between fermion and the O(3) vector order parameter $\mathbf{n}$ is denoted by $g_\psi$. Inside the antiferromagnetically ordered phase Dirac fermions possess excitation or mass gap $2 g_\psi \langle \bar{\psi} \mathbb{1} \otimes \boldsymbol \eta \cdot \mathbf{n} \psi \rangle$. The gradient expansion analysis (controlled by the mass gap) shows that a skyrmion acquires an induced chiral charge $Q_{5}=\langle \bar{\psi} \gamma_0 \gamma_5 \psi \rangle= 2 Q_{top}$, where $Q_{top}$ is the topological invariant or Pontryagin index for skyrmion configuration. Within the continuum description, the chiral charge acts as the generator of translational symmetry (an emergent U(1) symmetry when higher gradient kinetic terms are ignored). Inside the antiferromagnetically ordered phase, the skyrmion number and consequently the chiral charge $Q_5$ act as conserved quantities, thus freely mixing two bilinears $\bar{\psi}\hat{M}\psi$ and $\bar{\psi} \hat{M} \gamma_5 \psi$, where $[\hat{M}, \gamma_5]=0$, which cause hybridization between two inequivalent nodes. Consequently, skyrmion core supports translational symmetry breaking orders $\bar{\psi}\hat{M}\psi$ and $\bar{\psi} \hat{M} \gamma_5 \psi$ as fluctuating quantities. The specific choice $\hat{M}=\mathbb{1}$ corresponds to spin Peierls order, while other choices for $\hat{M}$ represent charge and current density wave orders. All of these singlet orders mix two valleys, and naturally break chiral or translational symmetry~\cite{FuSachdev,GoswamiSi2}. 

\subsection{Competition between Kondo singlets, spin Peierls and antiferromagnetic orders}
  
For the Kondo-Heisenberg model defined on the honeycomb lattice, we have to account for two species of eight-component fermions corresponding to conduction and f-electrons. Inside the antiferromagnetically ordered phase the low energy theory can be qualitatively understood in terms of the effective action
\begin{eqnarray}\label{S2}
S_2&=&\int d^2x d\tau \bar{\psi}[\gamma_\mu \partial_\mu + g_\psi \mathbb{1} \otimes \boldsymbol \eta \cdot \mathbf{n}]\psi \nonumber \\ &+& \int d^2x d\tau \bar{\chi}[\gamma_\mu \partial_\mu + g_\chi \mathbb{1} \otimes \boldsymbol \eta \cdot \mathbf{n}]\chi,   
\end{eqnarray}
where $\psi$ and $\chi$ capture two distinct eight-component Dirac fermions~\cite{GoswamiSi2,GoswamiSi3}. Crucially, the antiferrormagnetic sign of Kondo coupling is described by the condition $g_\psi g_\chi <0$ (same sign would represent Hund's coupling and describe spin-1 system). For simplicity all additional couplings between two species of fermions (residual quartic interactions) are being ignored. Both species of fermions give rise to induced chiral charges, while their sum vanishes. Interestingly, the difference between two types of induced chiral charge equals $4 Q_{top}$, i.e., $Q_{5,+}=\langle \bar{\psi} \gamma_0 \gamma_5 \psi \rangle+\langle \bar{\chi} \gamma_0 \gamma_5 \chi \rangle=0$ and $Q_{5,-}=\langle \bar{\psi} \gamma_0 \gamma_5 \psi \rangle - \langle \bar{\chi} \gamma_0 \gamma_5 \chi \rangle=4Q_{top}$. It has been shown that the relative chiral charge $Q_{5,-}$ (hence the skyrmion number) causes free rotation among several translational symmetry preserving Kondo singlet operators (mixing $\psi$ and $\chi$ at same valley) in addition to conventional translational symmetry breaking density wave operators. Therefore gradient expansion scheme provided important insight that the skyrmion texture supports several competing Kondo singlet operators, spin Peierls (bond density) as well as charge and current density wave orders inside the antiferromagnetic insulating phase~\cite{GoswamiSi2}. 

\subsection{One dimensional Kondo-Heisenberg model} 
A similar issue of interaction between Dirac fermions and topological defects of antiferromagnetic order has also been emphasized in one spatial dimension~\cite{Tsvelik,GoswamiSi1}. In one dimension the relevant topological defects are instantons or tunneling events for O(3) quantum nonlinear sigma model. However these instantons in two-dimensional Euclidean space, and static skyrmions of (2+1)-dimensional model have identical forms. By employing different field theoretic methods (direct gradient expansion and chiral anomaly), it has been found that the instanton number is directly related to the expectation value of bilinear $\bar{\psi} \gamma_5 \psi$ (which represents translational symmetry breaking, Ising spin-Peierls order). In the presence of Kondo coupling, one finds the competition between Kondo singlet formation and spin-Peierls order~\cite{GoswamiSi1}. This picture is also qualitatively supported by bosonization analysis.   

\subsection{Accomplishments of the present work}
However, the gradient expansion scheme only employs scattered states of Dirac fermions, while completely ignoring the effects of low energy bound states. How do these nonperturbative eigenstates affect the predictions of gradient expansion? Which are the most dominant singlet orders which can be nucleated after the antiferromagnetic order is destroyed by quantum fluctuations, causing a collapse of skyrmion excitation gap? In the present work we answer these important physical and technical questions. We first solve for the exact fermion eigenfunctions in the presence of topologically nontrivial skyrmion background to establish the induced chiral charge of skyrmion texture. Subsequently by employing these nonperturbative eigenstates, we evaluate the susceptibilities of different competing orders. Based on the susceptibilities, we demonstrate spin Peierls to be the most dominant translational symmetry breaking singlet order, which strongly competes against the static Kondo singlet formation. We also substantiate our results obtained in the continuum limit by calculations performed with lattice regularizations. Intriguingly, we find remarkable agreement between the analysis of this work and the predictions of perturbative field theory~\cite{GoswamiSi2} and more recent nonperturbative analysis of hedgehog-fermion interactions inside the paramagnetic phase~\cite{GoswamiSi3}. 

Since the two dimensional skyrmion texture describes the instanton or tunneling event of nonlinear sigma model in one spatial dimension, our methodology can be directly applied to the one dimensional problem (1+1-dimensional space-time) for computing the fermion determinant in the presence of topologically nontrivial dynamic background (it is equivalent to solving a fictitious two dimensional Hamiltonian defined in Euclidean space). Therefore, we can also extract the dynamic information regarding destruction of algebraic spin liquid in favor of competing Kondo singlet and spin Peierls phases for one dimensional Kondo-Heisenberg chain. Similarly, our methodology can be applied for many two and also three dimensional systems, supporting competition between spin-triplet and spin-singlet orders. 

The rest of the paper is organized as follows. We introduce the microscopic model and its continuum limit in Sec.~\ref{sec2}, and briefly discuss the results of gradient expansion in Sec.~\ref{sec3}. The calculations of nonperturbative eigenstates of Dirac fermions and order parameter susceptibilities are presented in Sec.~\ref{sec4}. The results from continuum limit are justified with lattice based calculations in Sec.~\ref{sec5}. We discuss the broader implications of our analysis in Sec.~\ref{sec6}, while we summarize our findings in Sec.~\ref{sec7}. Some details regarding the coupling between Dirac fermion and nonlinear sigma model fields, and chiral charge calculations are respectively relegated to Appendix~\ref{newsec} and Appendix~\ref{sec8}.

\section{Kondo lattice model on honeycomb lattice} \label{sec2}
\begin{figure}[hbtp]
\centering

\includegraphics[scale=0.23]{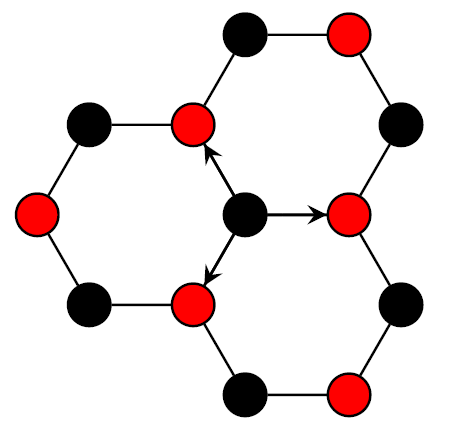}
\caption{The structure of honeycomb lattice, where the red and the black circles, respectively, denote two interpenetrating triangular sublattices A and B. Coordinate vectors $\boldsymbol{\delta}_i$ are shown as solid line with arrows.}\label{fig1}
\end{figure}

The Hamiltonian for Kondo-Heisenberg model on a honeycomb lattice is given by
\begin{widetext}
\begin{equation}\label{microH}
\begin{aligned}
&H =\sum_{\boldsymbol{r}_i\in A}\sum_{j=1}^3\bigl[-t_c\:  c^{\dagger}_{A,\alpha}\left(\boldsymbol{r}_i\right)c_{B,\alpha}\left(\boldsymbol{r}_i+\boldsymbol{\delta}_j\right)+h.c+J_H\:\boldsymbol{S}_A\left(\boldsymbol{r}_i\right)\cdot\boldsymbol{S}_B\left(\boldsymbol{r}_i+\boldsymbol{\delta}_j\right)+J_K\: c^{\dagger}_{A,\alpha}\left(\boldsymbol{r}_i\right)\frac{\boldsymbol{\eta}_{\alpha\beta}}{2}c_{A,\beta}\left(\boldsymbol{r}_i\right)\cdot \boldsymbol{S}_A\left(\boldsymbol{r}_i\right)\\
&+\frac{J_K}{3}\:c^{\dagger}_{B,\alpha}\left(\boldsymbol{r}_i+\boldsymbol{\delta}_j\right)\frac{\boldsymbol{\eta}_{\alpha\beta}}{2}c_{B,\beta}\left(\boldsymbol{r}_i+\boldsymbol{\delta}_j\right)\cdot\boldsymbol{S}_B\left(\boldsymbol{r}_i+\boldsymbol{\delta}_j\right)\bigr] ,
\end{aligned}
\end{equation}
\end{widetext}
where $c^{\dagger}_{A/B,\alpha/\beta}$ is the conduction electron creation operator, and A, B denote two interpenetrating triangular sublattices, and Pauli matrices $\boldsymbol \eta$ operate on spin indices $\alpha$ and $\beta$, and $\boldsymbol{\delta}_j$ are three coordination vectors connecting two sublattices, as shown in Fig.~\ref{fig1}. The explicit form of these vectors are $\boldsymbol{\delta}_1=\left(-\frac{a}{2},\frac{\sqrt{3}a}{2}\right)$, $\boldsymbol{\delta}_2=\left(a,0\right)$ and $\boldsymbol{\delta}_3=\left(-\frac{a}{2},-\frac{\sqrt{3}a}{2}\right)$, where $a$ is the lattice spacing. The local moments on  sublattice A and B are represented by $\boldsymbol{S}_A\left(\boldsymbol{r}_i\right)$ and $\boldsymbol{S}_B\left(\boldsymbol{r}_i+\boldsymbol{\delta}_j\right)$, respectively.  The RKKY coupling between local moment is modeled by nearest neighbor Heisenberg interaction with strength $J_H$, and $J_K$ is the Kondo coupling between conduction electron and local
  moment. We will consider both $J_H$ and $J_K$ to be antiferromagnetic, i.e., $J_H>0$ and $J_K>0$.

After linearizing the dispersion relation for fermions around two inequivalent nodal points of the hexagonal Brillouin zone (located at $\mathbf{K}_{\pm}$) and analytically continuing real time to imaginary time by setting $\tau=it$, the low energy effective physics of free conduction electron can be described by the imaginary time action:
\begin{equation}\label{S0}
S_0=\int dx^2 d\tau \overline{\psi}_{\alpha}\left(\gamma_0\otimes\eta_0\partial_t+v_{\psi}\gamma_j\otimes\eta_0\partial_j\right)\psi_{\alpha} ,
\end{equation}
where $v_{\psi}=\frac{\sqrt{3}t_c a}{3}$ is the Fermi velocity, spinor $\psi^T_{\alpha}=\left(c_{+,A,\alpha},c_{+,B,\alpha},c_{-,B,\alpha},c_{-,A,\alpha}\right)$, $\overline{\psi}_{\alpha}=\psi_{\alpha}\gamma_0$, $\pm$ is index for two valleys $K_{\pm}$, and $\alpha$ is spin index. The gamma matrices are defined as:
\begin{equation}\label{pauli}
\begin{aligned}
&\gamma_0=\tau_1\otimes \sigma_0=\begin{pmatrix}
0 & \sigma_0\\ \sigma_0 & 0
\end{pmatrix}, \gamma_j=i\tau_2\otimes\sigma_j=\begin{pmatrix}
0 & \sigma_j\\ -\sigma_j & 0
\end{pmatrix},\\
&\gamma_5=\tau_3\otimes\sigma_0=\begin{pmatrix}
\sigma_0& 0\\ 0& -\sigma_0
\end{pmatrix}
\end{aligned}
\end{equation}
where the Pauli matrices $\boldsymbol \sigma$, $\boldsymbol \tau$ respectively operate on the sublattice and valley indices.

Inside the antiferromagnetically ordered phase, the low energy physics of local moments can be described by QNL$\sigma$M~\cite{Duncan,MurthySachdev,Read,Chakravarty}:
\begin{equation}\label{Sn}
S_n=\frac{1}{2cg}\int d^2x d\tau \left[c^2\left(\partial_x\boldsymbol{n}\right)^2+\left(\partial_{\tau}\boldsymbol{n}\right)^2 \right]+iS_B\left[\boldsymbol{n}\right]
\end{equation}
The coupling constant $g$ has the dimension of length, and the antiferromagnetically ordered phase exists for $g$ smaller than a critical strength $g_c$~\cite{Chakravarty}. The last term $S_B\left[\boldsymbol{n}\right]$ corresponds to Berry phase, which vanishes inside the ordered phase. The Berry phase can be finite inside the paramagnetic phase, but it does not possess a simple continuum limit in (2+1) dimensions~\cite{Read}.

Now we incorporate the Kondo coupling, which captures the scattering between conduction electron spinor $\psi$ and the QNL$\sigma$M field $\boldsymbol{n}$ representing the local moment:
\begin{equation}\label{SK}
S_{K}=g_{K}\int d^2xd\tau\overline{\psi}_{\alpha}\gamma_3\boldsymbol{n}\cdot\boldsymbol{\eta}_{\alpha\beta}\psi_{\beta}
\end{equation}
Therefore, the low energy theory of antiferromagnetic phase for the Kondo-Heisenberg model can be described by: \begin{equation}\label{S}
S=S_0+S_n+S_K
\end{equation}

The lack of continuum representation for Berry's phase in (2+1)-dimensions makes it hard to analyze its consequence inside the paramagnetic phase based on the coarse grained representation. However, this can be circumvented by introducing auxiliary f-fermions for describing the local moments~\cite{GoswamiSi2}. We assume that the auxiliary f-fermions only hop to the nearest neighbor sites like the conduction fermions, with a hopping strength $t_f$. At low energies, these f-fermions can also be described by the Dirac equation with a new spinor $\chi^T_{\alpha}=\left(f_{+,A,\alpha},f_{+,B,\alpha},f_{-,B,\alpha},f_{-,A,\alpha}\right)$. Thus, the resulting low energy effective action for f-fermion inside AF phase is:
\begin{equation}\label{Sf}
S_f=\int dx^2 d\tau \overline{\chi}_{\alpha}\left[\gamma_0\otimes\eta_0\partial_t+v_{\chi}\gamma_j\otimes\eta_0\partial_j+g_{\chi}\gamma_3\boldsymbol{n}\cdot\boldsymbol{\eta}\right]_{\alpha\beta}\chi_{\beta}
\end{equation}
where $v_{\chi}=\frac{\sqrt{3}t_f a}{2}$. In fact, after integrating out the f-fermion degrees of freedom, this action will return to the same form of QNL$\sigma$M of Eq.~(\ref{Sn})~\cite{Abanov,FisherSenthil,TanakaHu}. We again remind the reader that the Berry phase vanishes inside the antiferromagnetically ordered phase and only becomes important for addressing the nature of paramagnetic phase. The Hamiltonian operator from Eq.~(\ref{SK}) involving only f-electrons would be:
\begin{equation}\label{Hf}
H_f=\tau_3\left[-i v_{\chi}\left(\sigma_1 \partial_1+\sigma_2 \partial_2\right)+g_{\chi}\boldsymbol{n}\cdot\boldsymbol{\eta}\sigma_3\right].
\end{equation}
Usually the introduction of auxiliary fermion description requires the introduction of Lagrange multiplier or constraint gauge fields. Since in this work we would be dealing with confined phases of matter such as antiferromagnet, spin Peierls or Kondo singlets, the constraint gauge field does not affect any of our conclusions regarding the competing order. For this reason we follow Ref.~\onlinecite{AffleckHaldane} and use an alternative method that avoids introduction of any constraint gauge fields. Within this method one considers actual $f$ electrons in the presence of sufficiently strong Hubbard interaction, which gives rise to an antiferromagnetic phase. The relevant steps are described in the Appendix~\ref{newsec}. 

Therefore, the Hamiltonian operator for the combined problem described by $S=S_0+S_f+S_K$ is given by
\begin{widetext}
\begin{equation}\label{HPsi}
H_{\Psi}=\tau_3\left[-iv_+\left(\sigma_1 \partial_1+\sigma_2 \partial_2\right)-iv_-\left(\sigma_1 \partial_1+\sigma_2 \partial_2\right)\rho_3+g_+\boldsymbol{n}\cdot\boldsymbol{\eta}\sigma_3+g_-\boldsymbol{n}\cdot\boldsymbol{\eta}\sigma_3\rho_3\right],
\end{equation}
\end{widetext}
which operates on the spinor $\Psi=\left(\psi,\chi\right)=\left(c_{A\alpha +},c_{B\alpha +},c_{B\alpha -},c_{A\alpha -},f_{A\alpha +},f_{B\alpha +},f_{B\alpha -},f_{A\alpha -}\right)$
where $v_{\pm}=\frac{v_c\pm v_f}{2}$ and $g_{\pm}=\frac{g_{K}\pm g_{\chi}}{2}$, and new Pauli matrices $\rho_i$ act on the flavor index representing conduction and f-electrons, $\left(\psi,\chi\right)$. Inside AF phase, we expect that the staggered magnetic moments of conduction electron $\psi$ and f-electron $\chi$ anti-align to each other. Therefore, we have $g_{K}g_{\chi}<0$~\cite{GoswamiSi2} . 

\section{Skyrmion, induced chiral charge and competing orders: perturbative argument}\label{sec3}
\begin{figure}[hbtp]
\centering
\includegraphics[scale=0.4]{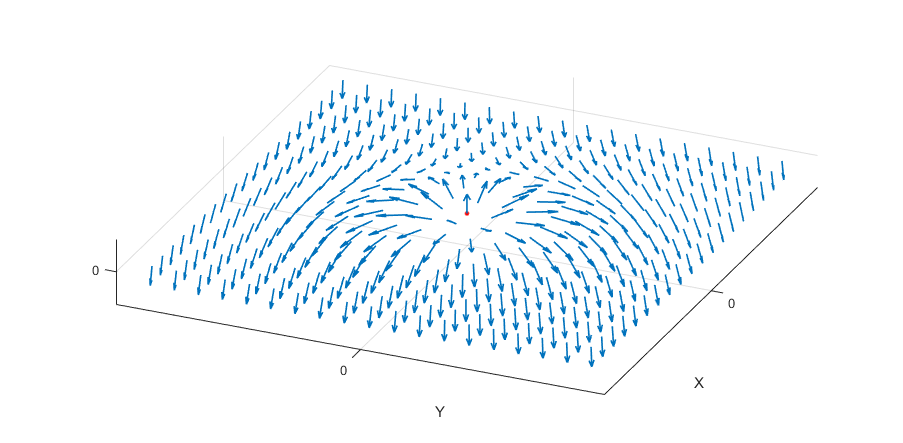}
\caption{Illustration of single skyrmion. The red dot denotes the origin of skyrmion core, and blue arrow is the direction of staggered magnetization or antiferromagnetic order parameter $\boldsymbol{n}$.}\label{fig2}
\end{figure}

The static nonsingular topological defect of QNL$\sigma$M in $2+1$ dimensions is called skyrmion, which satisfies the boundary condition $\boldsymbol{n}\left(r\rightarrow\infty\right)=\boldsymbol{n}^0$, where $r=\sqrt{x^2+y^2}$ and $\boldsymbol{n}^0$ is a constant unit vector. Therefore, the two-dimensional space is compactified onto a two sphere $S^2$ and the skyrmion configurations are defined by an integer topological charge also known as skyrmion number, since the homotopy group $\Pi_2\left(S^2\right)=Z$.
The skyrmion with topological charge $Q_{top}\in Z$ can have arbitrary profile function, provided it satisfies the boundary condition and the requirement that $\frac{1}{4\pi}\int d^2 x\boldsymbol{n}\cdot\partial_1\boldsymbol{n}\times\partial_2\boldsymbol{n}=Q_{top}$. Fig.~\ref{fig2} illustrates a real space profile for single skyrmion with $Q_{top}=1$.

\begin{figure}[hbtp]
\centering
\includegraphics[scale=0.32]{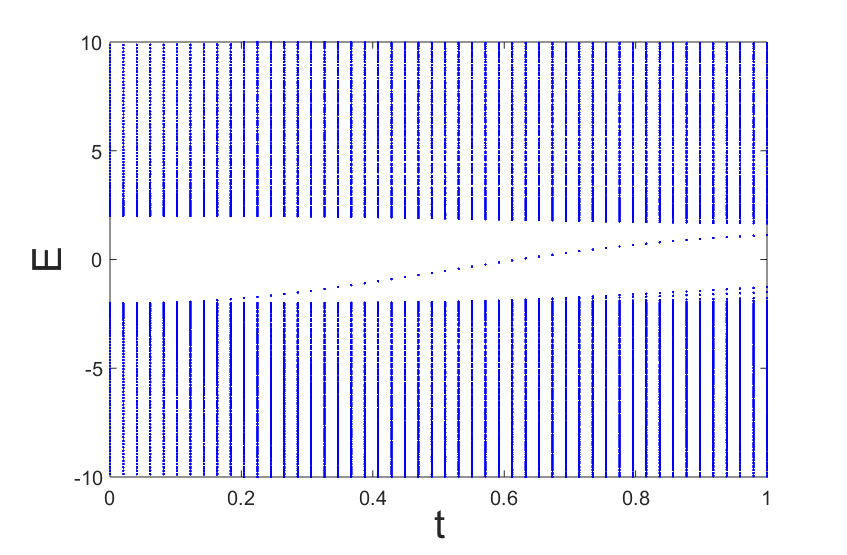}
\caption{The spectral flow for $+$ valley during the adiabatic formation of skyrmion. Here we choose coupling constant $g_{\chi}=2$ }\label{fig3}
\end{figure}

It is well known, when Dirac fermions are coupled to QNL$\sigma$M, the skyrmion textures will acquire induced fermion number~\cite{Jaroszewicz,Wilczek2,Carena,Abanov}. For Hamiltonian of Eq.~(\ref{Hf}), due to the overall matrix $\tau_3$ (appearing odd number of times) operating on two inequivalent valleys, the total induced fermionic charge vanishes. But the chiral charge, defined as the difference of fermion densities at two valleys, will be proportional to the topological charge of skyrmion:
\begin{equation}\label{Qpm}
\begin{aligned}
&Q_{\pm}\equiv \int d^2 x\langle \normOrd{f^{\dagger}_{\pm}f_{\pm}}\rangle=\pm \sign \left(g_{\chi}\right) Q_{top}, \\
&Q_{5}\equiv \int d^2 x\langle \normOrd{\chi^{\dagger}\tau_3 \chi}\rangle =  \int d^2 x\left(\langle \normOrd{f^{\dagger}_{+}f_{+}}\rangle- \langle \normOrd{f^{\dagger}_{-}f_{-}}\rangle\right), \\
&=2\sign\left(g_{\chi}\right)Q_{top},
\end{aligned}
\end{equation}
$Q_{\pm}$ are the charges for $\pm$ valleys, and $\normOrd{\,}$ denotes normal ordering operation. These relations can be proven by gradient expansion method~\cite{Abanov}, and the detailed derivation is provided in Appendix~\ref{sec8}. We can also verify this result numerically by solving for the spectral flow during adiabatic formation of skyrmion, as shown in Fig.~\ref{fig3}. We can simulate the formation of single (anti)skyrmion without loss of generality by assuming 
\begin{equation}
\boldsymbol{n}\left(\vec{r},t\right)=\left(\sin tf\left(r\right)\cos\theta,\sin tf\left(r\right)\sin\theta,\cos tf\left(r\right)\right),
\end{equation} 
where $f\left(r\right)=\pi e^{-\frac{r}{2}}$. One can easily verify that $Q_{top}=0$ at $t=0$ and $Q_{top}=-1$ at $t=1$, and the definition of (anti)skyrmion does not depend on the precise form of profile function. For $+$ valley, as shown in Fig.~\ref{fig3}, we find there is precisely one state that crosses zero energy (flowing out of negative energy states or filled Dirac sea) during the formation of skyrmion. Therefore, the induced charge is $-1$, just as Eq.~(\ref{Qpm}) suggests. The relation between the induced fermionic chiral charge of the system and the topological charge of skyrmion is a consequence of index theorem~\cite{Shifman2011}. 

Since $g_{K}g_{\chi}<0$, the induced chiral charges for conduction and f-electrons have opposite signs [electron $Q_{5,\psi}=2\sign\left(g_K\right)Q_{top}$ and f-electron $Q_{5,\chi}=2\sign\left(g_{\chi}\right)Q_{top}$]. This means if one state for conduction fermion sinks into the Dirac sea, there will be a state for f-electrons which will emerge out of the Dirac sea. Therefore, the net chiral charge of two species vanishes. Nonetheless, the difference between two chiral charges is quantized:
\begin{equation}\label{Q-}
Q_{-,\Psi}\equiv\int d^2 x\langle \normOrd{\Psi^{\dagger}\rho_3\tau_3 \Psi}\rangle=Q_{5,\psi}-Q_{5,\chi}=4\sign\left(g_K\right)Q_{top}
\end{equation}

Inside the AF ordered phase, the tunneling events described by singular hedgehog and antihedgehog configurations (space-time singularities) are linearly confined, leading to the conservation of skyrmion number. When the AF order is gradually suppressed by quantum fluctuations, the spin stiffness of the sigma model and the skyrmion energy cost decrease. On the paramagnetic side, the skyrmions excitation energy vanishes, and all topologically distinct skyrmion configurations become energetically degenerate. Hence, the tunneling events between different skyrmion configurations become important for determining how ground state degeneracy is lifted. Since $Q_5$ and $Q_{-,\Psi}$ are proportional to the topological charge $Q_{top}$ [as in Eq.~(\ref{Qpm}) and Eq.~(\ref{Q-})], $Q_5$ and $Q_{-,\Psi}$ would also be changed via tunneling events. Thus $Q_5$ and $Q_{-,\Psi}$ would act as fast variables inside paramagnetic phase, and their conjugate operators will serve as the appropriate slow variables or competing order parameters~\cite{FuSachdev,GoswamiSi2}.
 
Based on this argument, for one species of Dirac fermions [e.g., for Hamiltonian $H_{f}$ of Eq.~(\ref{Hf})], the corresponding spin-singlet competing orders in the particle-hole channel are found to be \begin{equation}
Q_M=\chi^{\dagger}\hat{M}e^{i\phi \tau_3}\chi.
\end{equation}
Here $\hat{M}$ is a $4\times 4$ matrix operating on sublattice and valley indices, and there are five distinct order parameters $O_M$, which are conjugate to chiral charge operator $\hat{Q}_{5}=\chi^{\dagger}\tau_3\chi$ , i.e., $\left[\hat{Q}_{5},O_M\right]\propto \chi^{\dagger}\hat{M}e^{i\left(\frac{\pi}{2}+\phi\right) \tau_3}\chi$, as indicated in TABLE~\ref{table1}. The first two correspond to components of the valence bond solid(VBS), which is also called Kekule bond density wave order or spin Peierls order breaking the translation symmetry unlike the usual AKLT state resulting from spin-1 model, and the final three correspond to different kinds of charge or current density wave orders~\cite{FuSachdev,GoswamiSi2}. However, only the components of VBS order anticommute with the whole Hamiltonian operator $H_f$ of Eq.~(\ref{Hf}), thus maximizing the energy gap inside the skyrmion core. Therefore, from a weak coupling perspective, the VBS order should be the most dominant competing order of antiferromagnetism.

 \begin{table}
\begin{tabular}{|p{1.9cm}||p{3.6cm}|p{2.9cm}|  }
 \hline
 Competing order & Matrix form $\hat{M}$ & anticommute with $H_{\Psi}$?  \\
 \hline
 Valence bond solid   & $\tau_1,\tau_2$ & Yes\\
  \hline
 Charge density wave   & $\tau_1\sigma_1,\tau_1\sigma_2$ & No\\
  \hline
 Current density wave   & $\tau_1\sigma_3$ & No\\
  \hline
\end{tabular}
  \caption{Competing orders for one species of fermion coupled to antiferromagnetic order parameter.}\label{table1}
 \end{table}

For the Kondo-Heisenberg model with two species of eight component Dirac fermions [see Eq.~(\ref{HPsi})], $Q_{-,\Psi}$ is proportional to the skyrmion number. Therefore, the conjugate operators of $Q_{-,\Psi}$ would serve as competing orders in the presence of antiferromagnetic Kondo coupling, and they are listed in TABLE~\ref{table2}. Besides VBS, charge and current density orders already found in TABLE~\ref{table1}, the presence of $\rho_3$ in $\hat{Q}_{-,\Psi}$ gives rise to additional competing orders involving $\rho_1$ or $\rho_2$, corresponding to hybridization of two species or Kondo singlet formation~\cite{GoswamiSi2}. While the VBS orders (with $\tau_1$,  $\tau_2$) always anticommute with the combined Hamiltonian, the Kondo singlet operators do not generically anticommute with the combined Hamiltonian. Hence from the weak coupling perspective, they may not be dominant competing orders inside the skyrmion core. Only for some special choice of parameters, some Kondo si
 nglet operators can anticommute with the effective Hamiltonian. Therefore, the gradient-expansion based results may not always predict the correct competing orders. In the following section, we circumvent this shortcoming of gradient-expansion scheme, by evaluating the exact eigenstates of Dirac Hamiltonian and subsequently computing the susceptibilities of different competing orders.

 \begin{table}
\begin{tabular}{|p{1.9cm}||p{3.6cm}|p{2.9cm}|  }
 \hline
 Competing order& Matrix form $\hat{M}$ & anticommute with $H_{\Psi}$?  \\
 \hline
 Valence bond solid   & $\tau_1,\tau_2$ & Yes\\
  \hline
 Charge density wave   & $\tau_1\sigma_1,\tau_1\sigma_2$ & No\\
  \hline
 Current density wave   & $\tau_1\sigma_3$ & No\\
  \hline
 Kondo singlet&   $\rho_1,\rho_2,\tau_3\rho_1,\tau_3\rho_2$  & Yes iff  $v_+=0$ and $g_+=0$   \\
  \hline
 Kondo singlet&$\sigma_3\rho_1,\sigma_3\rho_2,\tau_3\sigma_3\rho_1,\tau_3\sigma_3\rho_2$& Yes iff  $v_-=0$ and $g_+=0$\\
 \hline
\end{tabular}
  \caption{Competing spin-singlet orders in the presence of Kondo coupling}\label{table2}
 \end{table}

\section{Beyond pertubative argument} \label{sec4}

The eigenstates of Dirac fermions in the presence of skyrmion configurations of O(3) nonlinear sigma model have been previously discussed in Ref.~\onlinecite{Carena}. The main goal was to establish the induced fermion number due to spectral flow. But, the physical role of fermion doublers (present for any lattice model) and competing orders has not been addressed. By contrast, we would deal with fermion doublers arising from the underlying lattice model, and focus on identifying dominant competing orders residing in the skyrmion core. Therefore, we would compute susceptibilities of competing spin singlet order parameters, by using the exact eigenstates of Dirac fermions. This is a new development for the problem of interaction between Dirac fermions and O(3) skyrmion configurations.    
 
\subsection{Without Kondo coupling}

To calculate the local susceptibility of predicted competing orders in TABLE~\ref{table1}, we solve 
\begin{equation}\label{HfDelta}
\left(H_f+\Delta\hat{M}\right)\chi=E\chi
\end{equation}
on a finite disk of radius $R$ by performing exact diagonalization. We denote the Hamiltonian Eq.~\ref{Hf} with or without single skyrmion as $H_{f,S}$ and $H_{f,0}$, respectively. For single skyrmion, we choose the profile function of skyrmion $\boldsymbol{n}$ as
\begin{equation}\label{skyrmion}
\boldsymbol{n}=\left(\sin f\left(r\right)\cos\theta,\sin f\left(r\right)\sin\theta,\cos f\left(r\right)\right)
\end{equation}
where $\:f\left(r\right)=\pi e^{-\frac{r}{\lambda}}$ and $\lambda$ is the length scale for skyrmion. One can easily verify that in this case we have $\frac{1}{4\pi}\int d^2x \boldsymbol{n}\cdot\partial_1\boldsymbol{n}\times\partial_2\boldsymbol{n}=-1$.

The eigenstates of $H_{f,0}$ constitute a suitable basis for performing exact diagonalization. We choose the background field $\boldsymbol{n}=\left(0,0,1\right)$, such that: 
\begin{equation}\label{Hf0}
H_{f,0}=\tau_3\left[v_{\chi}\left(\sigma_1 k_1+\sigma_2 k_2\right)+g_{\chi}\eta_3\sigma_3\right].
\end{equation}
Since Hamiltonian $H_{f,0}$ commutes with grand spin operator $\hat{M}_3=-i\partial_{\theta}+\frac{\sigma_3}{2}+\frac{\eta_3}{2}$, $H_{f,0}$ and $\hat{M}_3$ can be simultaneoulsy diagonalized. The solutions for $H_{f,0}\chi=E\chi$ with fixed grand spin $m$ consist of the following linearly independent states:

\begin{equation}\label{basis}
\begin{aligned}
&\chi_{+,m,j,n,\eta=1} (r,\theta)=e^{im\theta}\begin{bmatrix} C_{\eta=1}\frac{v_{\chi}k_{m,j}}{n E_{m,j}-g_{\chi}} J_{m-1}\left(k_{m,j}r\right) e^{-i\theta}\\iC_{\eta=1}J_{m}\left(k_{m,j}r\right) \\0_{6\times1}\end{bmatrix}\\
&\chi_{+,m,j,n,\eta=-1} (r,\theta)=e^{im\theta}\begin{bmatrix} 0_{2\times 1}\\ C_{\eta=-1}\frac{v_{\chi}k_{m,j}}{n E_{m,j}+g_{\chi}} J_{m}\left(k_{m,j}r\right)\\iC_{\eta=-1}J_{m+1}\left(k_{m,j}r\right) e^{+i\theta}\\0_{4\times 1}\end{bmatrix}\\
&\chi_{-,m,j,n,\eta=1} (r,\theta)=e^{im\theta}\begin{bmatrix} 0_{4\times 1}\\ C_{\eta=1}\frac{v_{\chi}k_{m,j}}{n E_{m,j}-g_{\chi}} J_{m-1}\left(k_{m,j}r\right) e^{-i\theta}\\ iC_{\eta=1}J_{m}\left(k_{m,j}r\right) \\ 0_{2\times 1}\\\end{bmatrix}\\
&\chi_{-,m,j,n,\eta=-1} (r,\theta)=e^{im\theta}\begin{bmatrix} 0_{6\times 1 }\\ C_{\eta=-1}\frac{v_{\chi}k_{m,j}}{n E_{m,j}+g_{\chi}} J_{m}\left(k_{m,j}r\right)\\iC_{\eta=-1}J_{m+1}\left(k_{m,j}r\right) e^{+i\theta}\end{bmatrix}\\
\end{aligned}
\end{equation}
where the fist index $\pm$ means the $\pm$ valley, index $n=\pm 1$ denotes the solution with energy $\pm E_{m,j}$ , and $\eta$ is the spin index. We have chosen the boundary condition $J_m\left(k_{m,j}R\right)=0$, with index $j$ denoting the $j$-th zero of Bessel function $J_m\left(r\right)$ , so that the momentum $k_{m,j}$ and energy $E_{m,j}=\sqrt{v_{\chi}^2k^2_{m,j}+g^2_{\chi}}$ are quantized. The coefficient $C_{\eta}$ is determined by the normalization condition $\Vert \chi_{\pm,m,j,n,\eta}\Vert ^2\equiv\int d^2x \langle\chi_{\pm,m,j,n,\eta}\vert \chi_{\pm,m,j,n,\eta}\rangle=1$. 

\begin{figure}[t!]
\centering
\includegraphics[scale=0.31]{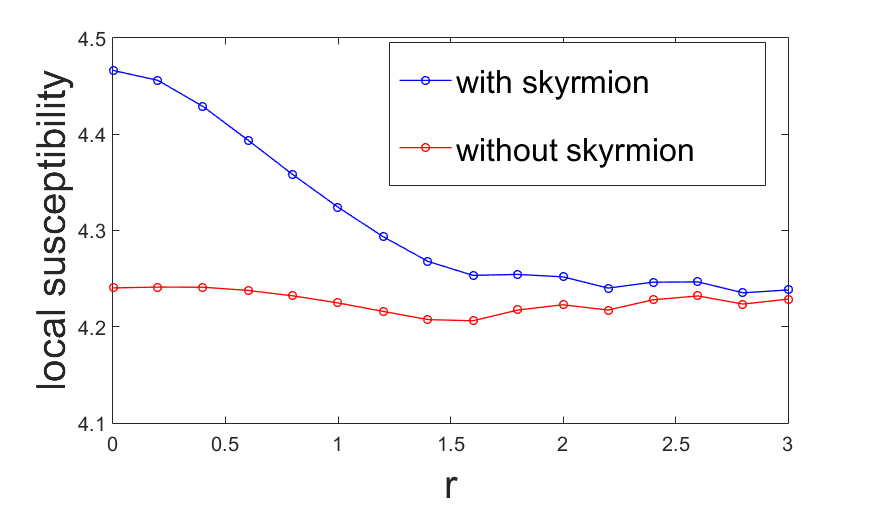}
\caption{The local susceptibility of VBS order $\hat{M}=\tau_1,\tau_2$ versus radial distance. The blue line and red line corresponds to the presence and the absence of skyrmion on origin, respectively. Once the skyrmion is present, the susceptibility of VBS order will gain obvious enhancement near the core of skyrmion defect.}
\label{susvbs}
\end{figure}

We then solve the equation Eq.~(\ref{HfDelta}) by diagonalizing the matrix with elements $\int d^2x \langle \chi_{\pm,m,j,n,\eta}\vert H_f+\Delta\hat{M}\vert \chi_{\pm,m',j',n',\eta}\rangle$. Besides the real space cut-off (i.e, the radius of disk $R$), we also impose a large momentum cut-off $\Lambda$, and the large grand spin cut-off $\overline{M}$. We choose our basis set spanning from grand spin  $-\overline{M}$ to $\overline{M}$. For $\hat{M}=\tau_1,\tau_2,\tau_1\sigma_3$, the Hamiltonian $H_{f}+\Delta\hat{M}$ commutes with the grand spin $\hat{M}_3$. Therefore, we can diagonalize the matrix in diagonal block with fixed value of grand spin $m$. While for $\hat{M}=\tau_1\sigma_1,\tau_1\sigma_2$ (which do not commute with $\hat{M}_3$), we have block off-diagonal elements and need to diagonalize the whole matrix at once.  

\begin{figure}[t!]
\centering
\includegraphics[scale=0.31]{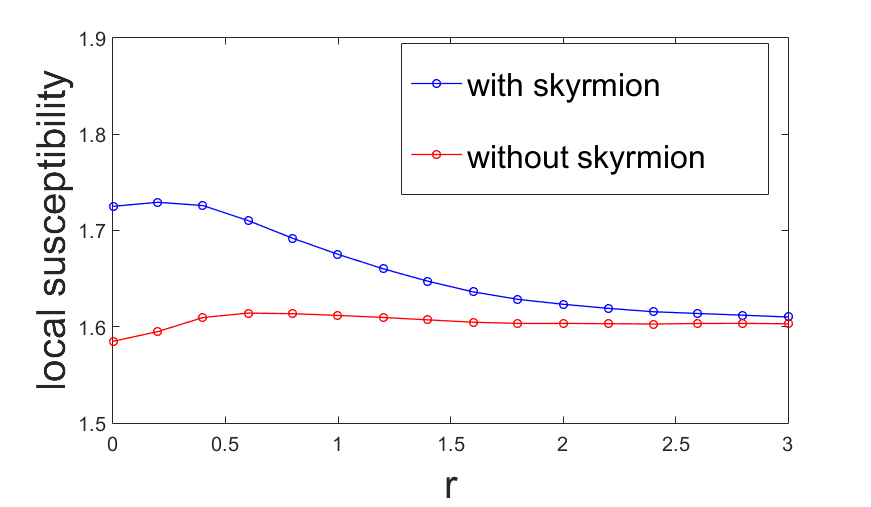}
\caption{The local susceptibility of charge density wave order $\hat{M}=\tau_1\sigma_1,\tau_1\sigma_2$ versus radial distance. Now the presence of skyrmion can still enhance the susceptibility, but the amount is smaller than VBS order.}
\label{suscdw}
\end{figure}

After finding the solutions for Eq.~(\ref{HfDelta}), we compute the local susceptibility for each candidate competing order by using \begin{equation}\label{localsus}
\chi_{M}\left(r\right)=\lim_{\Delta\rightarrow 0}\frac{\vert\langle\chi^{\dagger}\hat{M}\chi\rangle\vert}{\Delta}.
\end{equation}
 The local susceptibility diverges with momentum cut-off $\Lambda$ in two dimensions, but once we choose a finite momentum cut-off $\Lambda$, it converges with radius of disk $R$ and the maximum of grand spin $\overline{M}$. In this paper, we choose $R=8$, $\Lambda=8$, $\overline{M}=30$, the length scale of skyrmion $\lambda=2$, and the coupling constant $g_{\chi}=2$. 

\begin{figure}[b!]
\centering
\includegraphics[scale=0.31]{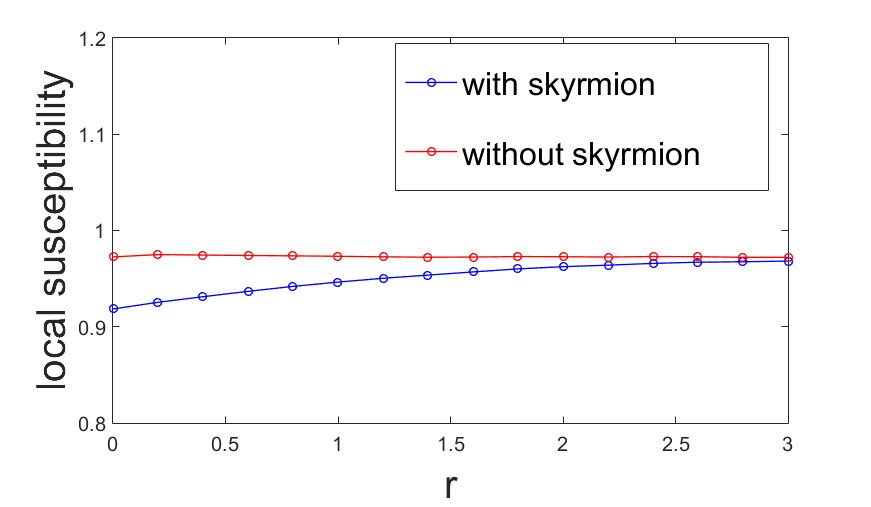}
\caption{The local susceptibility of current density wave order $\hat{M}=\tau_1\sigma_3$ versus radial distance. Instead of enhancement, the presence of skyrmion now suppresses the susceptibility of $\hat{M}=\tau_1\sigma_3$ near the core of skyrmion.}
\label{suscurrent}
\end{figure}

We have found that the local susceptibilities for VBS orders $\hat{M}=\tau_1$ or $\tau_2$ gain expected enhancement near the core of skyrmion, as shown in Fig.~\ref{susvbs}~\cite{footnote} 
On the other hand, for other candidate competing orders like charge density wave (with $ \tau_1\sigma_1$ and $ \tau_1\sigma_2$), the enhancement is less prominent, as shown in Fig.~\ref{suscdw}. Moreover, for current density wave $\tau_1\sigma_3$, the presence of skyrmion even suppresses the susceptibility, like Fig.~\ref{suscurrent}. The suppression of the susceptibility for current density wave $\tau_1\sigma_3$ demonstrates that the pertubative arguments of gradient-expansion scheme are not always sufficient.

\subsection{With Kondo coupling}

In the presence of Kondo coupling, we have to account two types of fermion fields $\psi$ and $\chi$, and the pertubative argument predicts that the VBS and Kondo singlet orders are important competing orders of antiferromagnetism [see TABLE~\ref{table2}]. We want to establish the validity of this prediction by using exact eigenstates of Dirac Hamiltonian. This is particularly important, since Kondo singlet operators do not generically anticommute with the Hamiltonian, and within the weak coupling picture fully anticommuting VBS would seem to be the dominant competing order. Whether the Kondo singlet orders can be favored over fully anticommuting VBS over a wide range of microscopic parameter regime is not clear from the weak coupling arguments. By contrast, our physical intuition suggests that the Kondo singlets should be stabilized over a finite parameter region~\cite{Si_PhysicaB2006}. Here we address this issue by solving the eigenstates of 
\begin{equation}\label{HPDelta}
\left(H_{\Psi}+\Delta\hat{M}\right)\Psi=E\Psi,
\end{equation}
for each $\hat{M}$ identified in TABLE~\ref{table2}, and then computing the local susceptibility by employing
\begin{equation}\label{localsus2}
\chi_{M}\left(r\right)=\lim_{\Delta\rightarrow 0}\frac{\vert\langle\Psi^{\dagger}\hat{M}\Psi\rangle\vert}{\Delta}.
\end{equation}

\begin{figure}[b!]
\centering
\includegraphics[scale=0.31]{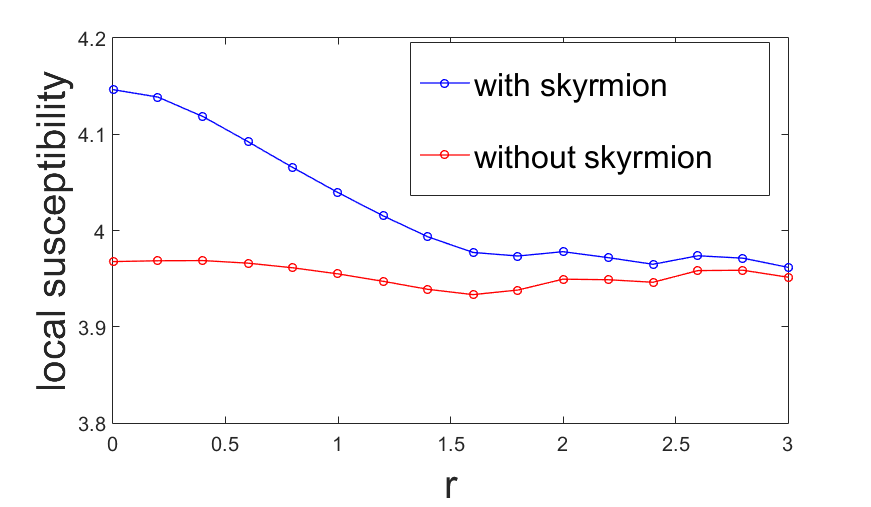}
\caption{The local susceptibility of Kondo singlet orders $\hat{M}=\rho_1,\rho_2,\tau_3\rho_1,\tau_3\rho_2$ ($\hat{M}=\sigma_3\rho_1,\sigma_3\rho_2,\sigma_3\tau_3\rho_1,\sigma_3\tau_3\rho_2$) with $v_{\psi}=1$, $v_{\chi}=-1$, $g_{\psi}=2$, and $g_{\chi}=-3$ ($v_{\psi}=1$, $v_{\chi}=1$, $g_{\psi}=2$, and $g_{\chi}=-3$). The enhancement of susceptibility of these Kondo singlet order by skyrmion can still sustain obviously, even with parameters beyond where perturbative argument can be applied. The enhancement of susceptibility of Kondo orders in fact can sustain to very broad parameter space.}
\label{suskon}
\end{figure}

For diagonalizing this Hamiltonian we use the basis set:  $\Psi_{\pm,m,j,n,\eta}=\left(\psi_{\pm,m,j,n,\eta},\chi_{\pm,m,j,n,\eta}\right)$, where $\chi_{\pm,m,j,n,\eta}$'s have been already defined in Eq.~(\ref{basis}) and $\psi_{\pm,m,j,n,\eta}$'s are defined as:

\begin{equation}\label{basis2}
\begin{aligned}
&\psi_{+,m,j,n,\eta=1} (r,\theta)=e^{im\theta}\begin{bmatrix} D_{\eta=1}\frac{v_{\psi}k_{m,j}}{n E'_{m,j}-g_{K}} J_{m-1}\left(k_{m,j}r\right) e^{-i\theta}\\ iD_{\eta=1}J_{m}\left(k_{m,j}r\right) \\0_{6\times 1}\end{bmatrix}\\
&\psi_{+,m,j,n,\eta=-1} (r,\theta)=e^{im\theta}\begin{bmatrix} 0_{2\times 1} \\ D_{\eta=-1}\frac{v_{\psi}k_{m,j}}{n E'_{m,j}+g_{K}} J_{m}\left(k_{m,j}r\right)\\iD_{\eta=-1}J_{m+1}\left(k_{m,j}r\right) e^{+i\theta}\\0_{4\times 1}\end{bmatrix}\\
&\psi_{-,m,j,n,\eta=1} (r,\theta)=e^{im\theta}\begin{bmatrix} 0_{4\times 1}\\ D_{\eta=1}\frac{v_{\psi}k_{m,j}}{n E'_{m,j}-g_{K}} J_{m-1}\left(k_{m,j}r\right) e^{-i\theta}\\ iD_{\eta=1}J_{m}\left(k_{m,j}r\right) \\ 0_{2\times 1}\\\end{bmatrix}\\
&\psi_{-,m,j,n,\eta=-1} (r,\theta)=e^{im\theta}\begin{bmatrix} 0_{6\times 1} \\ D_{\eta=-1}\frac{v_{\psi}k_{m,j}}{n E'_{m,j}+g_{K}} J_{m}\left(k_{m,j}r\right)\\iD_{\eta=-1}J_{m+1}\left(k_{m,j}r\right) e^{+i\theta}\end{bmatrix}.\\
\end{aligned}
\end{equation}
Here  $E'_{m,j}=\sqrt{v_{\psi}^2k^2_{m,j}+g^2_{\psi}}$ and the coefficient $D_{\eta}$ is obtained from the normalization condition $\Vert \psi_{\pm,m,j,n,\eta}\Vert ^2\equiv\int d^2x \langle\psi_{\pm,m,j,n,\eta}\vert\psi_{\pm,m,j,n,\eta}\rangle=1$. We diagonalize the matrix with elements $\int d^2 x\langle \Psi_{\pm,m,j,n,\eta}\vert H_{\Psi}+\Delta\hat{M}\vert \Psi_{\pm,m',j',n',\eta}\rangle$.

As shown in Fig.~\ref{suskon}, we have found that the enhancement of these Kondo singlet orders by skyrmion is comparable with the enhancement of VBS orders. Moreover, the enhancement of Kondo singlets is also sustained over a broad parameter space, including the regime where perturbative arguments may not be applicable. Therefore, we conclude that the Kondo singlet and VBS orders act as the dominant competing orders inside a skyrmion core. Therefore, the paramagnetic phase in the global phase diagram can support both of these competing singlet orders.

 \begin{figure}[b!]
\centering
\includegraphics[scale=0.31]{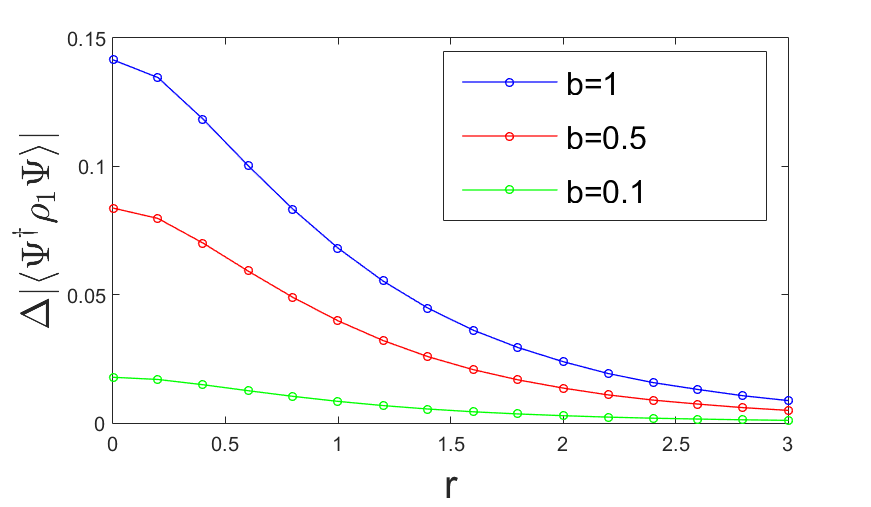}
\caption{The difference of Kondo singlet order $\langle \rho_1\rangle$ between the presence and the absence of the skyrmion on origin for different fluctuation strength $b$. We choose $v_{\psi}=1$, $v_{\chi=}=-1$, $g_{\psi}=2$, and $g_{\chi}=-3$ and $Q=0$ here. $\Delta\vert\langle\Psi^{\dagger}\rho_1\Psi\rangle\vert$ is defined as $\Delta\vert\langle\Psi^{\dagger}\rho_1\Psi\rangle\vert\equiv\vert \langle \Psi^{\dagger}\rho_1\Psi\rangle\vert_{1}-\vert \langle \Psi^{\dagger}\rho_1\Psi\rangle\vert_{0}$, where $\vert\langle \Psi^{\dagger}\rho_1\Psi\rangle\vert_{1}$ and $\vert\langle \Psi^{\dagger}\rho_1\Psi\rangle\vert_{0}$ means $\vert\langle \Psi^{\dagger}\rho_1\Psi\rangle\vert$ calculated in the background with and without single skyrmion, respectively.}
\label{crossover}
\end{figure}

 \subsection{Crossover between VBS and Kondo order}
The Hamiltonian of Eq.~(\ref{HPsi}) is only useful for describing low energy physics inside the antiferromagnetic phase. In the vicinity of a magnet to paramagnet phase transition, such description is not sufficient to capture all features of the Kondo lattice model, since the fluctuations for competing channels and residual interactions in those channels can become important. The effective Hamiltonian describing the competition among VBS, Kondo singlet, and AF phases for a Kondo lattice model can be postulated to have the form  
\begin{equation}\label{Hfluc}
H=H_{\Psi}+b\rho_1+Q\tau_1\frac{\rho_0-\rho_3}{2} ,
\end{equation}
where $b$ and  $Q$ capture the fluctuations for Kondo and VBS channels and increase with $J_K$, $J_H$ respectively~\cite{Saremi,Pixley}. The presence of $\frac{\rho_0-\rho_3}{2}$ reflects that the VBS order in a Kondo lattice model can only be generated through the frustrated RKKY interactions between local moments.

  \begin{figure}[t!]
\centering




\includegraphics[scale=0.23]{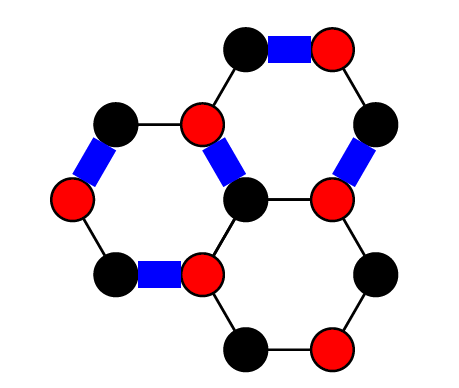}
\caption{The VBS pattern with  $ \delta t_{\boldsymbol{r}_i,\boldsymbol{r}_i+\boldsymbol{\delta}_j}=\Delta e^{i\boldsymbol{K}_+\cdot\boldsymbol{\delta}_j}e^{i\boldsymbol{G}\cdot\boldsymbol{r}_i}/3+\text{H.c}$ . The blue thick(black thin) lines indicate hopping amplitude is increased(decreased) by $\frac{2\Delta}{3}$($\frac{\Delta}{3}$).}
\label{vbspattern}
\end{figure}

From the perspective of AF Hamiltonian $H_{\Psi}$, the fluctuations of VBS and Kondo channels serve as external perturbation, and thus induce the corresponding order parameters approximately as:
\begin{equation}\label{order}
\begin{aligned}
&\langle \chi^{\dagger}\tau_1\chi\rangle\cong Q\chi_{VBS}\left(r\right)\\
&\langle \Psi^{\dagger}\rho_1\Psi\rangle\cong b\chi_{Kondo}\left(r\right)\\
\end{aligned}
\end{equation}
where $\chi_{VBS}\left(r\right)$ and $\chi_{Kondo}\left(r\right)$ are the local susceptibilities of VBS and Kondo orders, respectively. 

Since we have already observed that the skyrmion defect of AF order can enhance the susceptibility of VBS and Kondo order inside AF phase, we expect that the VBS and Kondo order parameters induced by these fluctuations will also be enhanced by skyrmion. Moreover, once the $J_K$($J_H$) is enlarged, that is, the fluctuation into Kondo(VBS) channel is larger, from Eq.~(\ref{order}), the resulting enhancement of Kondo(VBS) order parameter by skyrmion should also be enlarged.

This behavior actually is also manifested by solving the Hamiltonian of Eq.~(\ref{Hfluc}) directly and computing the resulting order parameters, as shown in Fig.~\ref{crossover}. Therefore, once we increase the Kondo coupling in microscopic Kondo lattice model, the skyrmion will eventually favor Kondo order over the VBS order, causing the transition from VBS to Kondo phases. This result gives us a unifying point of view to understand the crossover between VBS and Kondo orders in a Kondo lattice model, beginning from the antiferromagnetic phase.

  \begin{figure}[t!]
\centering
\includegraphics[scale=0.24]{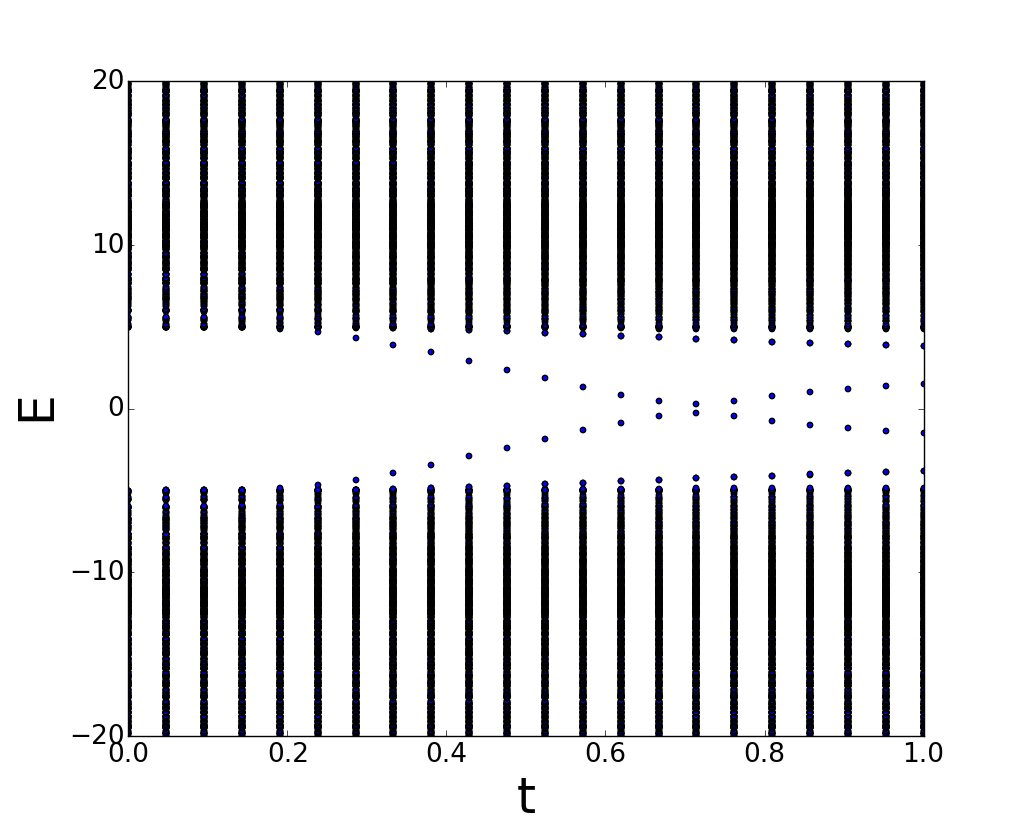}
\caption{The spectrum flow during the formation of skyrmion for lattice Hamiltonian ~\ref{Hlattice} involving f-electron only. We choose coupling constant $J_H=5$, $t_f=10$ and simulate the formation of skyrmion by $\boldsymbol{n}\left(\vec{r}_i,t\right)=\left(\sin tf\left(r_i\right)\cos\theta,\sin tf\left(r_i\right)\sin\theta,\cos tf\left(r_i\right)\right)$, where $f\left(r_i\right)=\pi e^{-\frac{r_i}{2}}$ and $r_i$ is the radial position of the site $i$. There is one state flowing from negative state to positive state, and precisely one state flowing oppositely. This is just a reflection of relation ~\ref{Qpm}, since the spectrum here consists of + and - valley.}
\label{laspecflow}
\end{figure}

\section{Justification by lattice models} \label{sec5}
\begin{widetext}

\begin{figure} [t!]
\centering
\hspace{0.5 in}
\subfigure[VBS order $\hat{M}=\tau_1$]{
\includegraphics[scale=0.21]{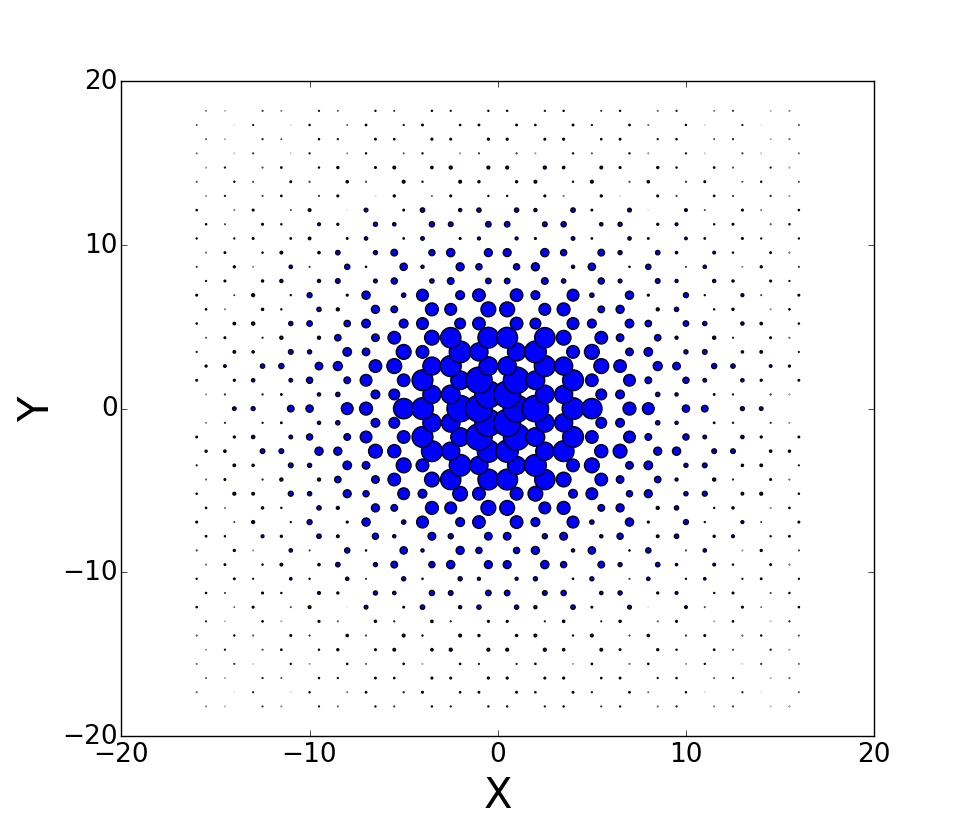}}
\hspace{0.5 in}
\subfigure[Charge density wave order $\hat{M}=\tau_1\sigma_1$]{
\includegraphics[scale=0.21]{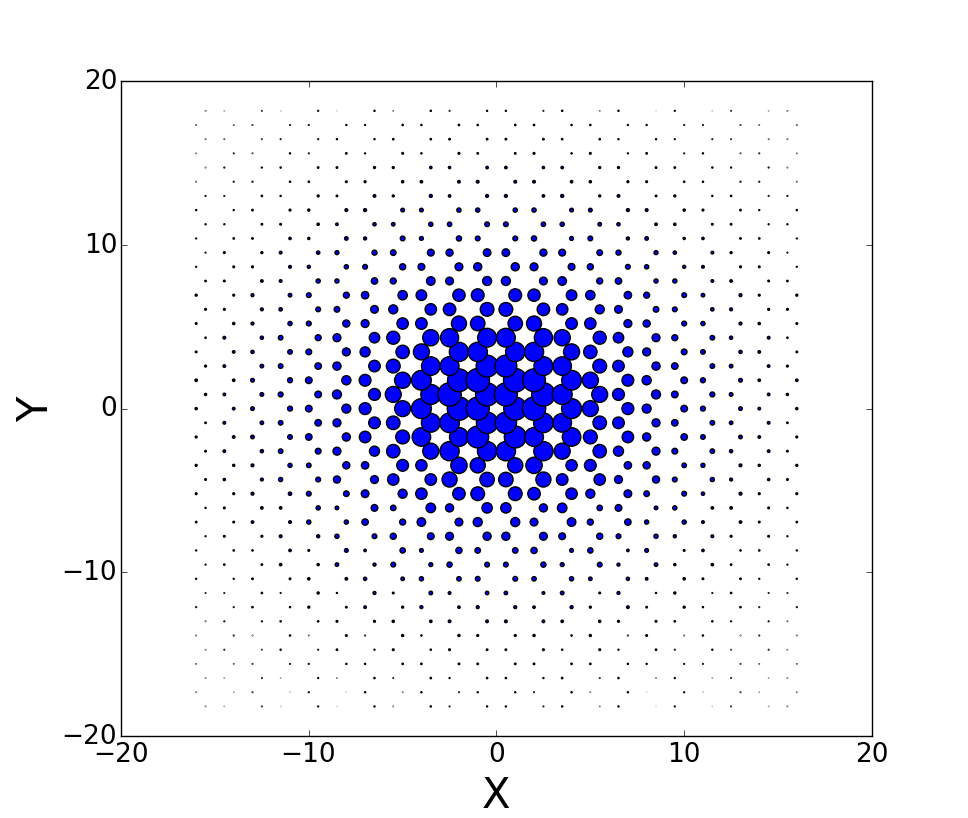}}
\subfigure[Current density wave order  $\hat{M}=\tau_1\sigma_3$]{
\hspace{0.5 in}
\includegraphics[scale=0.21]{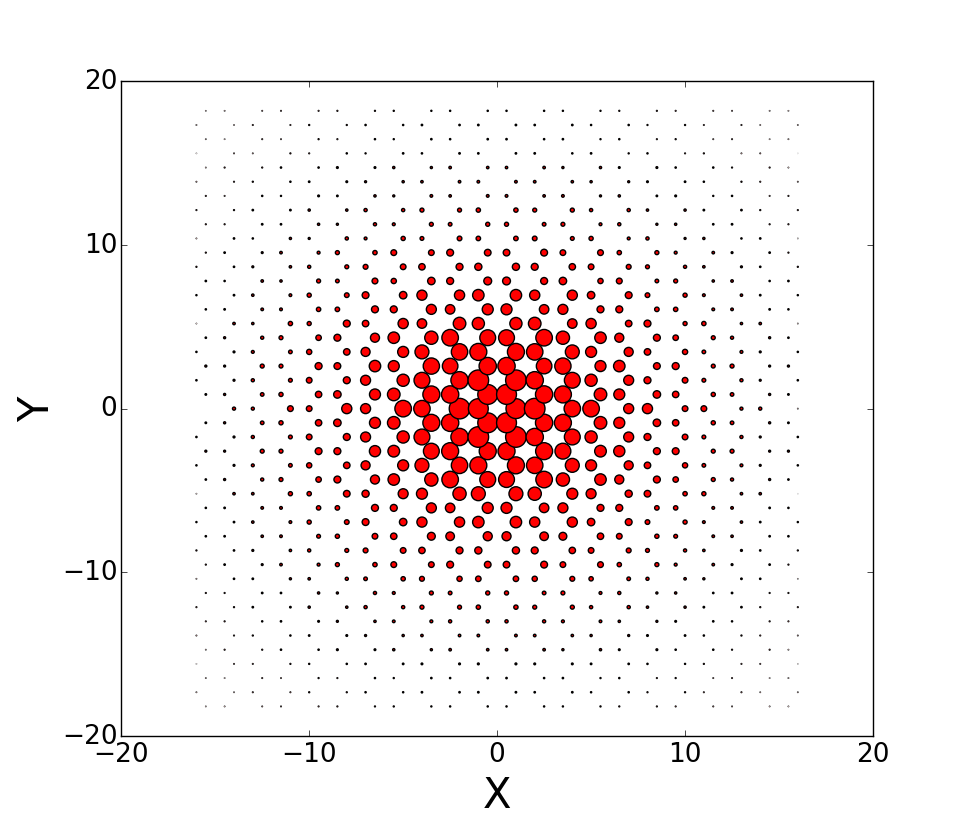}}
\subfigure[Kondo singlet order  $\hat{M}=\rho_1$]{
\hspace{0.5 in}
\includegraphics[scale=0.21]{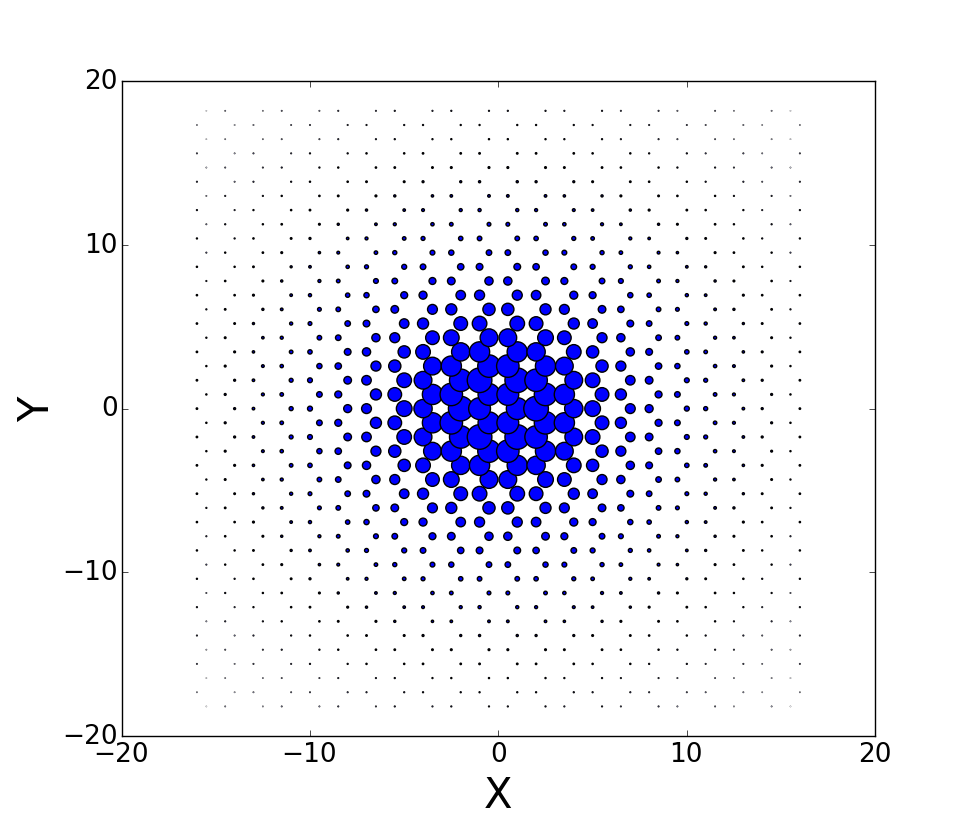}}
\caption{The difference between distinct order parameters in the presence and absence of a skyrmion at the origin. Here we solve the lattice Hamiltonian whose low energy effective model is $H_{\Psi}+\Delta\hat{M}$. The blue circle means the difference is positive, and the bigger circle indicates the difference is larger. We can see obviously that VBS, charge density wave, and Kondo singlet order gain enhancement near the core of skyrmion, while the current density wave is suppressed due to the presence of skyrmion, which is consistent with the results from low energy Hamiltonian in last section. Similar conclusions hold for other order parameters listed in TABLE~\ref{table1} and ~\ref{table2}.}
\label{fig:comparison} 
\label{orderdiff}
\end{figure}
\end{widetext}

So far, the model we relied on are different kinds of low energy effective Dirac-type Hamiltonian. In these models, the presence of large momentum cut-off is practically inevitable, even though all of our conclusions hold regardless of cut-offs. In order to further justify these results, we have also solved the lattices models in the presence of skyrmion defect (whose low energy effective theory is equivalent to $H_{\Psi}+\Delta\hat{M}$ for different candidate competing orders in TABLE~\ref{table1} and Table~\ref{table2}) through exact diagonalization. For example, the VBS order $\hat{M}=\tau_1$ can be generated through the lattice model:
\begin{widetext}
\begin{equation}\label{Hlattice}
H=\sum_{\langle ij\rangle\alpha}\left[-t_f f^{\dagger}_{i,\alpha} f_{j,\alpha}- t_c c^{\dagger}_{i,\alpha} c_{j,\alpha}+h.c\right]+\sum_{i\alpha\beta} \left[J_H\left(-1\right)^{A=0,B=1}f^{\dagger}_{i,\alpha}\frac{\left(\boldsymbol{n}\cdot\boldsymbol{\eta}\right)_{\alpha\beta}}{2} f_{i,\beta}+ J_K\left(-1\right)^{A=1,B=0}c^{\dagger}_{i,\alpha}\frac{\left(\boldsymbol{n}\cdot\boldsymbol{\eta}\right)_{\alpha\beta}}{2} c_{i,\beta}\right]
\end{equation}
\end{widetext}
if we replace $t_f\rightarrow t_f+\delta t_{\boldsymbol{r}_i,\boldsymbol{r}_i+\boldsymbol{\delta}_j}$ and $t_c\rightarrow t_c+\delta t_{\boldsymbol{r}_i,\boldsymbol{r}_i+\boldsymbol{\delta}_j}$, where $ \delta t_{\boldsymbol{r}_i,\boldsymbol{r}_i+\boldsymbol{\delta}_j}=\Delta e^{i\boldsymbol{K}_+\cdot\boldsymbol{\delta}_j}e^{i\boldsymbol{G}\cdot\boldsymbol{r}_i}/3+\text{H.c}$ and $\boldsymbol{G}=\boldsymbol{K}_+-\boldsymbol{K}_-$, as Fig.~\ref{vbspattern}. The resulting low energy effective Hamiltonian of this model is exactly the same as $H_{\Psi}+\Delta\tau_1$ ~\cite{Changyu}. By assigning the skyrmion configuration for local moment field $\boldsymbol{n}$, we can explore its influence on VBS order parameter in a lattice model. The presence of skyrmion in the lattice model also causes spectral flow events as in Fig.~\ref{laspecflow}, which is consistent with the low energy continuum theory. The VBS order parameter in lattice model can be extracted through the nearest-neighbor hopping amplitude.  After solving the lattice model, we can see that the presence of skyrmion enhances the VBS order parameter as shown in Fig.~\ref{orderdiff}(a). Similar results for other competing orders are presented in
  Fig.~\ref{orderdiff}. All of the results are consistent with our previous findings based on low energy Dirac theory. 

\section{Discussion} \label{sec6}

In the field theory literature, the nonperturbative eigenstates of Dirac fermions have been already employed for computation of induced fermion numbers~\cite{JackiwRebbi1,THooft,JackiwRebbi2, Callias1,Callias2,Wilczek1,MacKenzie,Kahana,MaNieh,BBoyanovsky,SBoyanovsky,Carena,GoswamiSi3}. Some famous examples are the induced chiral charge of a domain wall in one dimension, and baryon number of O(4) skyrmions in three dimensions. A similar analysis has also been performed for O(3) skyrmions in two spatial dimensions. However, the physical issue of competing orders and the determination of dominant fluctuating order based on nonperturbative eigenbasis are new aspects of the present work. To the best of our knowledge previous analysis along this direction has been restricted
to competing orders in a vortex core (defects of Abelian theory). 

Since we are explicitly solving for the eigenstates of the Dirac Hamiltonian, we can also employ these states for computing the competing orders away 
from half-filling. Not much is known for such a situation from perturbative field theory. Th chiral charge also acts as the generators for translational 
symmetry breaking paired states (FFLO states). At half-filling they do not produce fully gapped states and are less favorable compared 
to spin-Peierls order (causing Dirac mass gap). However, when we move away from the special case of half-filling, the paired states are more 
effective in gap formation. Therefore, we expect FFLO phases to become more important in the generic situation of finite carrier density. 
Even current and charge density wave orders which were earlier disfavored compared to spin-Peierls order can become more important 
(as none of them are able to effectively gap out the Fermi surface). Such intriguing competition among particle-hole and particle-particle 
channel condensates are germane to understanding the generic global phase diagrams of correlated metals, and will be elucidated 
in a future publication.

Our methodology can be easily adapted for both higher and lower dimensional problems. Specifically, the computed energy-eigenstates for two dimensional model in the presence of skyrmion configuration can be directly taken over as the complete eigenbasis for evaluating the fermion determinant in one dimension in the presence of dynamic instanton background. Such calculations can again be performed both at and away from half-filled limit to unveil the competition among spin-Peierls, Kondo singlets and paired states, which have been suggested by different perturbative calculations as well as some density matrix renormalization group analysis.

\section{Conclusion} \label{sec7}
We
addressed
 the nonperturbative aspects of interaction between topological defects and fermions, and how it can give rise to competition among different order parameters. Specifically, we considered the interaction between topologically nontrivial skyrmion configurations of antiferromagnetic phase and fermionic quasiparticles in two spatial dimensions. 
 To make
  progress we have modeled the fermionic excitations by Dirac fermions. Beginning with a half-filled Kondo-Heisenberg model on a honeycomb lattice, we investigated fluctuating orders that can be supported by skyrmion core inside the antiferromagnetically ordered insulating phase. Inside this ordered state, we have considered the coupling between conduction and f-electrons to the collective mode, described in terms of an O(3) quantum nonlinear sigma model(QNL$\sigma$M). By employing perturbative field theory, exact numerical and analytical solutions for eigenfunctions of Dirac fermions in the presence of 
 a single skyrmion we have established the competition between magnetism, Kondo singlet formation and spin Peierls order. Our specific goal was to establish
 a framework for finding dominant order parameter, which can be
 adapted for many other problems involving the interaction between fermion and topological defects. The perturbative field theory calculation of Goldstone-Wilczek current for our model suggests the presence of several translational symmetry breaking orders such as charge, bond and current density waves as well as translational symmetry preserving Kondo singlet formation.
 However, this method does not clearly specify the dominant incipient order. Therefore, we have explicitly computed the susceptibilities for all possible local Dirac bilinears by using nonperturbatively determined eigenfunctions. Our analysis thus provides 
 strong evidence that the global phase diagram of Kondo-Heisenberg can support 
 a variety of competing singlet orders 
 from skyrmion condensation (violation of skyrmion number) on the paramagnetic side.
 All of our results from continuum model have been consistently justified by analysis performed on suitable lattice model. This general strategy for identifying dominant competing orders mediated by topological defects can be useful in both one and three spatial dimensions.

\appendix

\section{Coupling between fermions and nonlinear sigma model}\label{newsec}
Since we are working with a bipartite honeycomb lattice, an intraunit cell antiferromagnetic phase (N\'{e}el order) describes the ground state of a nearest neighbor Heisenberg model. The nonlinear sigma model description for this phase is usually derived by employing a large spin approximation. However, for describing the competing spin singlet orders such as spin Peierls and Kondo singlet it is more advantageous to work with a fermionic description. This is similar to the methods of Affleck and Haldane~\cite{AffleckHaldane} for one dimensional spin-1/2 chain. The antiferromagnetic phase for honeycomb lattice can only be obtained from a Hubbard model for sufficiently strong onsite repulsion, as the density of states for two dimensional Dirac fermion vanishes at zero energy. The repulsive Hubbard interaction, $H_{int}=U\sum_i \; n_{i,\uparrow} n_{i, \downarrow}$, where $n_{i,s}$ is the density operator for spin projection $s=\uparrow / \downarrow$, can be decoupled in the magnetic channel by performing the following Hubbard-Stratonovich transformation
\begin{widetext}
\begin{eqnarray}
 \int \; dc^\dagger_i dc_i \; \exp \left[-\int_{0}^{\frac{\hbar}{k_BT}} \frac{d\tau}{\hbar} H_{int}\right] &=& \int \; dc^\dagger_i dc_i d\mathbf{M}_i  \;  \exp \bigg[- \int_{0}^{\frac{\hbar}{k_BT}} \frac{d\tau}{\hbar} \bigg \{\sum_i \bigg[\frac{3 \mathbf{M}^2_i}{2 U}+ \mathbf{M}_i \cdot c^\dagger_{i,s} \boldsymbol \eta_{s,s^\prime} c_{i,s^\prime} \bigg] \nonumber \\ &&+ \frac{U}{2} \sum_{i,s} \; n_{i,s}  \bigg \} \bigg].\label{Hubbard1}
\end{eqnarray}
\end{widetext}
Notice that the Hubbard interaction has been decoupled in the magnetic channel in terms of the vectors $\mathbf{M}_i$ ($i=A, B$ are assigned to two sublattices), where $\mathbf{M}_i=(U/3)\langle c^\dagger_{i,s} \boldsymbol \eta_{s s^\prime} c_{i,s^\prime} \rangle$. In the process of mean-field decoupling the chemical potential has to be shifted by the amount $U/2$, to maintain the condition of half-filling. The antiferromagnetic phase corresponds to the choice $\mathbf{M}_A=-\mathbf{M}_B$. Due to the vanishing density of states the antiferromagnetic phase arises for $U>U_c$. Within the continuum limit this leads to the following effective action
\begin{eqnarray}
S=\int d^2x d\tau \left[\bar{\psi} \gamma_\mu \partial_\mu + g \bar{\psi} \mathbf{M}\cdot \boldsymbol \eta \psi + \frac{\mathbf{M}^2}{2}\right],\label{Hubbard2}
\end{eqnarray}
with $g \propto \sqrt{U}$. At $U=U_c$ (equivalently $g=g_c$), we have an itinerant version of paramagnetic semimetal to antiferromagnet quantum phase transition, where the fermion fields and both longitudinal and transverse parts of the order parameter constitute gapless or critical excitations. For $U>U_c$, the amplitude of the order parameter $|\mathbf{M}| \sim |U-U_c|^{\beta}$ is finite, and away from the itinerant critical regime i.e., at the length scales larger than the correlation length $\xi \sim |U-U_c|^{-\nu}$ we can effectively freeze the amplitude fluctuations of the magnetic order parameter. Since we can denote $\mathbf{M}=|\mathbf{M}| \; \mathbf{n}$, where $\mathbf{n}$ is the unit vector or nonlinear sigma model field, after freezing $\mathbf{M}|$ Eq.~(\ref{Hubbard2}) can be reduced to 
$$S=\int d^2x d\tau \left[\bar{\psi} \gamma_\mu \partial_\mu + g_\psi \bar{\psi} \mathbf{n}\cdot \boldsymbol \eta \psi \right].$$This allows us to work with a nonlinear sigma model coupled to Dirac fermions, as used in the main text. The longitudinal part of the nonlinear sigma model field gives rise to a charge gap for the Dirac fermions, and after integrating out the Dirac fermions by following~\cite{Abanov,FisherSenthil,TanakaHu} one can obtain a nonlinear sigma model. The ordered phase of the nonlinear sigma model indeed corresponds to the ordered phase obtained within the large spin approximation of nearest neighbor Heisenberg model. An advantage for the effective theory is that the bare stiffness for nonlinear sigma model does not guarantee a global long range order, and it remains possible that the emergent gapped/insulating phase supports a nonmagnetic competing order, where the Berry phase for the sigma model does not vanish and follows from the evaluation of fermion determinant~\cite{FisherSenthil,TanakaHu}.  

\section{Topological charge of skyrmion and induced charge} \label{sec8}
The induced charge for each valley is defined as the difference of charge in each valley between system with and without skyrmion (vacuum):
\begin{equation}\label{Qpm2}
\begin{aligned}
&Q_{\pm}=\int_{-\infty}^0 dE \rho_{S,\pm}(E)-\int_{-\infty}^0 dE  \rho_{0,\pm}(E)\\
&=-\frac{1}{2}\int_{-\infty}^{\infty}dE\rho_{S,\pm}(E)\rm \sign(E)=-\frac{1}{2}\eta_{\pm}\\
\end{aligned}
\end{equation}
where $\rho_{S,\pm}(E)$ and $\rho_{0,\pm}(E)$ is the density of state at energy $E$ with and without skyrmion for $\pm$ valley, respectively, $\eta_{\pm}=\int_{-\infty}^{\infty}dE\rho_{S,\pm}(E)\rm \sign(E)=\int_{0}^{\infty}dE\left(\rho_{S,\pm}(E)-\rho_{S,\pm}(-E)\right)$ is called the spectral asymmetry, and we have used the fact that system without skyrmion field has charge conjugate symmetry.

Since Hamiltonian of Eq.~(\ref{Hf}) does not break valley symmetry, it can be decoupled into each valley space as $H_{\pm}$, the density of states in each valley is well-defined as:
\begin{equation}\label{dos}
\rho_{\pm}(E)\equiv \Tr\delta\left(H_{\pm}-E\right)= \frac{1}{\pi}\im\Tr\left(\frac{1}{H_{\pm}-E-i\epsilon}\right)
\end{equation}

The spectral asymmetry then is:
\begin{equation}\label{specasy}
\begin{aligned}
&\eta_{\pm}=\int_{0}^{\infty}dE\left(\rho_{S,\pm}(E)-\rho_{S,\pm}(-E)\right)\\
&=\frac{1}{\pi}\int_{0}^{\infty}dE\left(\im\Tr\frac{1}{H_{\pm}-E-i\epsilon}-\im\Tr\frac{1}{H_{\pm}+E-i\epsilon}\right)\\
&=\frac{1}{\pi}\int_{0}^{\infty}dE\im\Tr\left(\frac{1}{H_{\pm}-E-i\epsilon}+\frac{1}{H_{\pm}+E+i\epsilon}\right)
\end{aligned}
\end{equation} 
where we use the identity $\frac{1}{x\pm i\eta}=P\left(\frac{1}{x}\right)\mp i\pi\delta(x)$. By changing the variable $z=E+i\epsilon$, we have 
\begin{equation}\label{specasy2}
\begin{aligned}
&\eta_{\pm}=\frac{1}{\pi}\im\int^{\infty+i\epsilon}_{i\epsilon} dz\Tr\left(\frac{1}{H_{\pm}-z}+\frac{1}{H_{\pm}+z}\right)\\
&=\frac{2}{\pi}\im\int^{\infty+i\epsilon}_{i\epsilon} dz\Tr\left(H_{\pm}\frac{1}{H_{\pm}^2-z^2}\right)
\end{aligned}
\end{equation}
In our case, $H_{\pm}=\pm\left(v_{\chi}\left(\sigma_1 k_1+\sigma_2 k_2\right)+g_{\chi}\boldsymbol{n}\cdot\boldsymbol{\eta}\sigma_3\right)=\pm\left(H_0+I\right)$, where $H_0=v_{\chi}\left(\sigma_1 k_1+\sigma_2 k_2\right)$ and $I=g_{\chi}\boldsymbol{n}\cdot\boldsymbol{\eta}\sigma_3$, thus
$H_{\pm}^2=H_0^2+V=-v_{\chi}^2\nabla^2+g_{\chi}^2-ig_{\chi}v_{\chi}\sigma_3\sigma^i\partial_i\boldsymbol{n}\cdot\boldsymbol{\eta}$, where $H_0^2=-v_{\chi}^2\nabla^2+g_{\chi}^2$ and $V=-ig_{\chi}v_{\chi}\sigma_3\sigma^i\partial_i\boldsymbol{n}\cdot\boldsymbol{\eta}$

We assume that background field varies very slowly compared with coupling constant, that is $\lvert \nabla\boldsymbol{n}\rvert \ll g_{\chi}$, and then expand $\eta_{\pm}$ in order of $\lvert \nabla\boldsymbol{n}\rvert/ g_{\chi}$:
\begin{equation}\label{expand}
\begin{aligned}
&\Tr\left(H_{\pm}\frac{1}{H_{\pm}^2-z^2}\right)=\Tr\left(H_{\pm}\frac{1}{H_0^2+V-z^2}\right)\\
&=\Tr\left(H_{\pm}G_0\left(z\right)\left(1+G_0\left(z\right)V\right)^{-1}\right)\\
&=\Tr\left(H_{\pm}G_0\left(z\right)\sum_{n=0}^{\infty}\left(-G_0\left(z\right)V\right)^n\right)
\end{aligned}
\end{equation}
where $G_0\left(z\right)=\frac{1}{H_0^2-z^2}$.

By identity $G_0\left(z\right)V=VG_0\left(z\right)+G_0\left(z\right)\left[V,H_0\right]G_0\left(z\right)$, we are now able to separate the trace into pure momentum and real space part, and then do the trace separately. The non-vanishing leading order of Eq.~(\ref{expand}) will be $\pm\Tr\left(IV^2\right)\Tr\left(G_0^3\left(z\right)\right)$. Since
\begin{equation}
\begin{aligned}
&\Tr\left(IV^2\right)=-g^3_{\chi}v^2_{\chi}\Tr\left(\sigma^3_3\boldsymbol{n}\cdot\boldsymbol{\eta}\sigma^i\partial_i\boldsymbol{n}\cdot\boldsymbol{\eta}\sigma^j\partial_j\boldsymbol{n}\cdot\boldsymbol{\eta}\right)\\
&=- g^3_{\chi}v^2_{\chi}\Tr\left(\sigma_3\sigma^i\sigma^j\eta^a \eta^b \eta^c n_a\partial_i n_b \partial_j n_c\right)\\
&=-4g^3_{\chi}v^2_{\chi}\int d^2x \epsilon^{ij} \epsilon^{abc} n_a\partial_i n_b \partial_j n_c \\
&\Tr\left(G_0^3\left(z\right)\right)=\Tr\left(\frac{1}{H_0^2-z^2}\right)^3=\Tr\left(\frac{1}{-v_{\chi}^2\nabla^2+g_{\chi}^2-z^2}\right)^3\\
&=\int \frac{d k }{\left(2\pi\right)^2} \frac{2\pi k}{\left(v_{\chi}^2k^2+g^2_{\chi}-z^2\right)^3}=\frac{1}{8\pi v_{\chi}^2\left(g_{\chi}^2-z^2\right)^2}
\end{aligned}
\end{equation}
Consequently, the leading order of $\eta_{\pm}$ will be 
\begin{equation}
\begin{aligned}
& \frac{\mp 8 g^3_{\chi}v^2_{\chi}}{\pi}\im\int^{\infty+i\epsilon}_{i\epsilon} \frac{dz}{8\pi v_{\chi}^2\left(g_{\chi}^2-z^2\right)^2} \int d^2x \epsilon^{ij} \epsilon^{abc} n_a\partial_i n_b \partial_j n_c \\
&= \frac{\mp \sign\left(g_{\chi}\right)}{4\pi}\int d^2x \epsilon^{ij} \epsilon^{abc} n_a\partial_i n_b \partial_j n_c
\end{aligned}
\end{equation}
Therefore, \begin{equation}
\begin{aligned}
&Q_{\pm}=-\frac{1}{2}\eta_{\pm}= \frac{ \pm \sign\left(g_{\chi}\right)}{8\pi}\int d^2x \epsilon^{ij} \epsilon^{abc} n_a\partial_i n_b \partial_j n_c\\
&=\pm\sign\left(g_{\chi}\right)Q_{top}
\end{aligned}
\end{equation}

\acknowledgements 
This work has been supported in part by the NSF grant DMR-1611392 and the Robert A. Welch Foundation Grant No. C-1411 (C.-C. L and Q.S.), 
and JQI-NSF-PFC and LPS-MPO-CMTC (P.G.). We acknowledge the hospitality of the Aspen Center for Physics (the NSF Grant No. PHY-1607611).
One of us (Q.S.) also acknowledges the hospitality of Department of Physics, University of California at Berkeley.


\begin{thebibliography}{}
\bibitem{footnote} Actually, local susceptibility of $\tau_1$ and $\tau_2$ are exactly the same, since $\tau_1$ and $\tau_2$ can be transformed to each other by transforming the basis and the Hamiltonian Eq.~(\ref{HfDelta}) is invariant under such basis transformation. The same conclusion holds for the charge density wave orders $\tau_1\sigma_1$ and $\tau_1\sigma_2$, and different Kondo singlet orders in next section.
\end{thebibliography}
\end{document}